\documentclass{iau}
\usepackage{graphicx}
\newcommand{\be}{\begin{equation}}
\newcommand{\ee}{\end{equation}}
\newcommand{\bary}{\begin{eqnarray}}
\newcommand{\eary}{\end{eqnarray}}
\newcommand{\en}{E_\nu}

\title[JD 11.~~Pre-solar grains and AGB stars] 
{Hadronic flares and associated neutrinos for Markarian 421}

\author[Antonio Marinelli, Barbara Patricelli \& Nissim Fraija]   
{Antonio Marinelli$^1$, Barbara Patricelli$^2$
 \and Nissim Fraija$^2$}

\affiliation{$^1$Instituto de F\'isica, Universidad Nacional Aut\'onoma de 
M\'exico, Circuito Exterior, C.U., A. Postal 70-264, 04510 M\'exico D.F., 
M\'exico. \\ email: {\tt antonio.marinelli@pi.infn.it, antonio.marinelli@fisica.unam.mx} \\[\affilskip]
$^2$Instituto de Astronom\' ia, Universidad Nacional Aut\'onoma de M\'exico, Circuito Exterior, 
C.U., A. Postal 70-264, 04510 M\'exico D.F., M\'exico. \\email: {\tt bpatricelli@astro.unam.mx, nifraija@astro.unam.mx - Luc Binette Foundation}}

\pubyear{2014}
\volume{313}  
\pagerange{119--126}
\jname{Extragalactic jet from every angle}
\editors{F.Massaro, C.C.Cheung, E.Lopez \& A.Siemiginowska, eds.}
\begin{document}

\maketitle

\begin{abstract}

Markarian 421 (Mrk 421) is one of the brightest, fastest and closest  BL Lac object known. Its very high energy (VHE) spectrum has been successfully modeled with both leptonic and hadronic models and not conclusive results have been achieved yet about the origin of its VHE emission. Here we investigate the possibility that a fraction of the VHE flares of Mrk 421 are due to hadronic processes and calculate the expected neutrino flux associated. We introduce the obtained neutrino flux in a Monte Carlo simulation to see the expectation for a Km$^{3}$ Cherenkov neutrino telescope. 
\end{abstract}

\firstsection 

\section{Limits on the fraction of time spent by Mrk 421 in hadronic flaring states}

Blazars spend a fraction of their time in a lowest steady flux level, called baseline state and another fraction in flaring states. The duty cycle is the percentage of time that the source spends in flaring states.  \cite[Abdo et al. (2014)]{2014ApJ...782..110A} presented an estimation of the duty cycle above 1 TeV ($DC_{TeV}$) of Mrk 421 for different flare flux thresholds and compared their results with the X-ray duty cycle $DC_X$ estimated by \cite[Resconi et al. (2009)]{2009A&A...502..499R}. If only the leptonic synchrotron self- Compton (SSC) process occurred, we should expect that $DC_X$ is similar to the $DC_{TeV}$, since in this framework  synchrotron X-ray photons are always up scattered by relativistic electrons to VHE. Instead, \cite[Abdo et al. (2014)]{2014ApJ...782..110A} have shown that, for flares with a flux above 3 standard deviations from the baseline flux ($F_{\rm baseline}$), $DC_{\rm TeV}$ is greater than $DC_X$, suggesting the possible presence of emission processes additional to the SSC, for instance hadronic processes. If we consider that both leptonic and hadronic processes are responsible for the TeV flares, we can write the $DC_{\rm TeV}$ as the sum of the leptonic and hadronic components ($\rm {DC}^{\rm lept}$ and DC$^{\rm hard}$):
\begin{equation}\label{eq:totDC}
DC_{\rm TeV}=\rm{DC}^{\rm lept}+\rm{DC}^{\rm hadr}=DC_X+\frac{T_{\rm flare}^{\rm hadr}}{T_{\rm obs}},
\end{equation}
where $T_{\rm obs}$ is the total observation period. Therefore, we can estimate $T_{\rm flare}^{\rm hadr}=(DC_{\rm TeV}-DC_{\rm X})\times T_{\rm obs}$.
To calculate $T_{\rm flare}^{\rm hadr}$ we consider $F_{\rm baseline}=0.33$ Crab, corresponding to the mean of the gaussian function describing the baseline state of Mrk 421 (\cite[Tluczykont et al. 2010]{2010A&A...524A..48T}). For this value of $F_{\rm baseline}$,  $DC_{\rm TeV}=(27 \pm 7) \%$ (\cite[Abdo et al. 2014]{2014ApJ...782..110A}), while \cite[Resconi et al. (2009)]{2009A&A...502..499R} estimated  $DC_X= (18.1 \pm 0.5) \%$. Therefore,  we have $T_{\rm flare}^{\rm hadr}=\sim9 \% \times T_{\rm obs}$.

\section{VHE Neutrino expectations and Results}
\begin{figure}
\begin{center}
\includegraphics[scale=0.5]{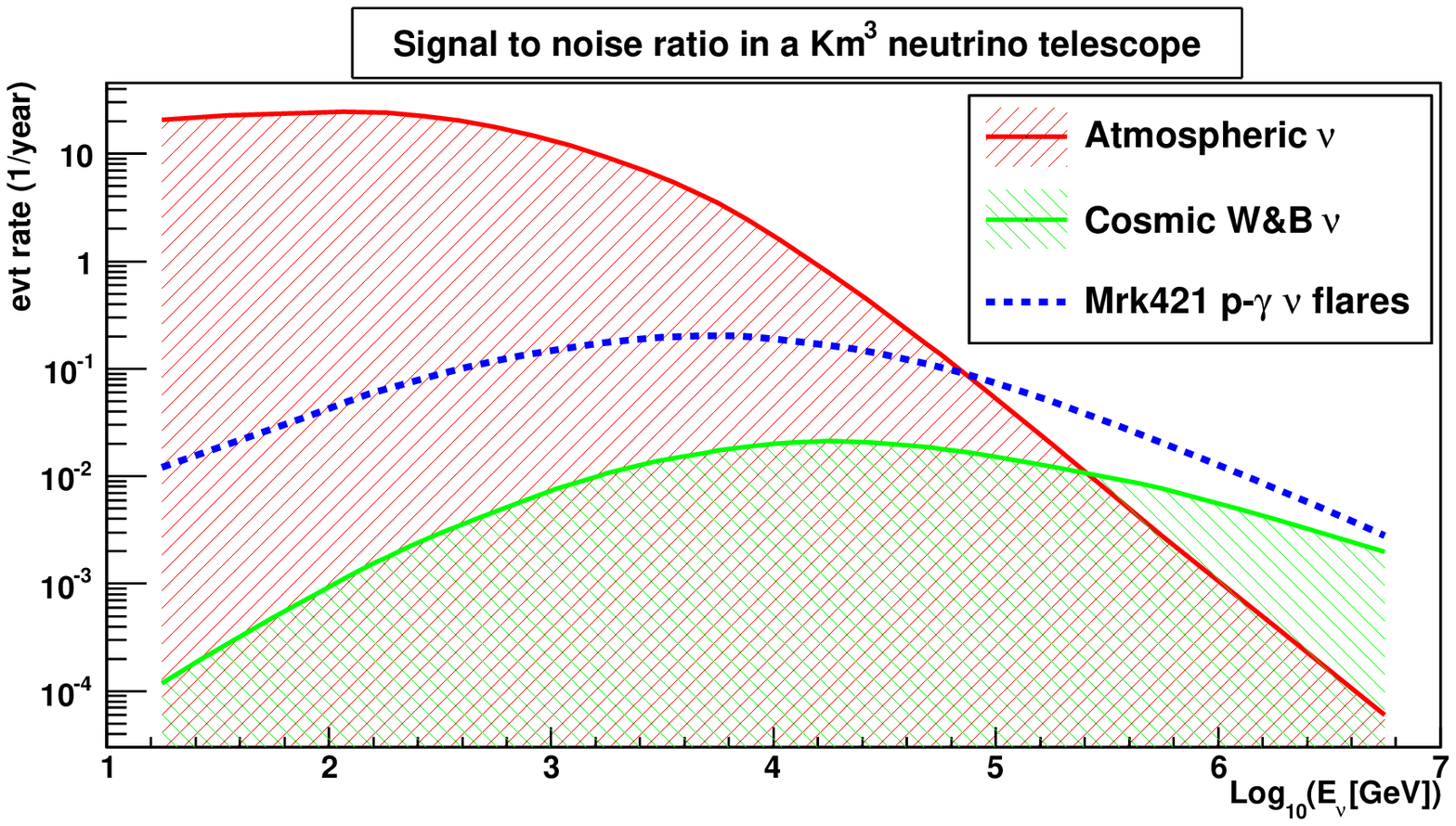}
\caption{In this plot the neutrino signal to noise ratio for  Mrk421 considering a Km$^{3}$ Cherenkov telescope. The atmospheric and cosmic neutrino ``backgrounds'' are obtained considering a circular region around Mrk421 of 1$^{\circ}.$}\label{sigtonoisenu}
\end{center}
\end{figure}

Assuming that the TeV gamma-ray spectrum of Mrk421 is produced by the interaction of Fermi-accelerated protons with SSC photons, a neutrino counterpart is expected.
In particular for each p$\gamma$ interaction we have gamma-ray production through $\pi^{\circ}\rightarrow\gamma\gamma$ and neutrino production through 
$\pi^{\pm}\rightarrow e^{\pm}+\nu_{\mu}/\bar{\nu}_{\mu}+\bar{\nu}_{\mu}/\nu_{\mu}+\nu_{e}/\bar{\nu}_{e}$. The spectral indices of neutrino and gamma-ray spectra are considered similar  $\alpha\simeq \alpha_\nu$ while the carried energy is slightly different: each neutrino brings 5$\%$  of the initial proton energy ($\en=1/20\,E_p$) while  each photon brings around 16.7$\%$.  With these considerations the normalization factor of neutrino spectrum $A_{(p\gamma,\nu)}$ is related to the normalization factor of gamma-ray spectrum $A_{(p\gamma,\gamma)}$ through $A_{(p\gamma,\nu)}=K\cdot A_{(p\gamma,\gamma)}\,\epsilon_0^{-2}\, (2)^{-\alpha+2}$.
Extending the spectrum of expected neutrino to maximum energies detectable by a Km$^{3}$ Cherenkov detector array, we can obtain the number expected neutrino events detected as:
\begin{equation}
N_{ev} \approx\,T \rho_{water/ice}\,N_A\,V_{eff}\,\int_{E_{min}}^{E_{max}}\sigma_{\nu}\,A_{(p\gamma,\nu)}\left(\frac{E_{\nu}}{TeV}\right)^{-\alpha}dE_{\nu}.
\label{nuMCevt}
\end{equation}
Where $N_A$ is the Avogadro number, $\rho_{water/ice}$ is the density of environment for the neutrino telescope, $E_{min}$ and $E_{max}$ are the low and high energy threshold considered,
$V_{eff}$ is the $\nu_{\mu}+\bar\nu_{\mu}$ effective volume, obtained through Monte Carlo simulation, for a hypothetical Km$^{3}$ neutrino telescope considering the neutrino source at the declination of Mrk421. In Fig. \ref{sigtonoisenu} we can see the signal to noise ratio for one year of data-taking. As a signal we consider the neutrinos obtained assuming the gamma-ray spectrum of Mrk421 Veritas high state (\cite[Acciari et al. 2001]{2011ApJ...738...25A}), taking into account the $T_{\rm flare}^{\rm hadr}$ estimated in the previous section, while the ``backgrounds'' are represented by the atmospheric and cosmic neutrinos reconstructed in $1^{\circ}$ around Mrk421. With a $T_{\rm flare}^{\rm hadr}=\sim9 \% \times T_{\rm obs}$ more than one decade of observation is needed for a Km$^{3}$ neutrino telescope like IceCube to see a neutrino track-event related to the hadronic component of Mrk421 high states.


\begin{thebibliography}{}

\bibitem[Abdo \etal\ (2014)]{2014ApJ...782..110A}
{Abdo, A.A., et al.} 2014
\textit{ApJ}, 782, 110

\bibitem[Acciari \etal\ (2011)]{2011ApJ...738...25A}
{Acciari, V.A., et al.} 2011
\textit{ApJ}, 738, 25

\bibitem[Resconi \etal\ (2009)]{2009A&A...502..499R}
{Resconi, E., et al.} 2009
\textit{A\&A}, 502, 499

\bibitem[Tluczykont \etal\ (2010)]{2010A&A...524A..48T}
{Tluczykont, M., et al.} 2010
\textit{A\&A}, 524, 48


\end{thebibliography}
\end{document}